\documentclass{article}
\usepackage{spconf,amsmath,graphicx,amsthm,amssymb,color,makecell,url,bm}
\usepackage{hyperref}

\title{DCTTS: Discrete Diffusion Model with Contrastive Learning for Text-to-Speech Generation }
\name{ Zhichao Wu, \qquad  Qiulin Li, \qquad  Sixing Liu, \qquad  Qun Yang \sthanks{ Corresponding author: qun.yang@nuaa.edu.cn}}
\address{Nanjing University of Aeronautics and Astronautics,Nanjing,China}
\begin{document}
%
\maketitle
\begin{abstract}
In the Text-to-speech(TTS) task, the latent diffusion model has excellent fidelity and generalization, but its expensive resource consumption and slow inference speed have always been a challenging. This paper proposes Discrete Diffusion Model with Contrastive Learning for Text-to-Speech Generation(DCTTS). The following contributions are made by DCTTS: 1) The TTS diffusion model based on discrete space significantly lowers the computational consumption of the diffusion model and improves sampling speed; 2) The contrastive learning method based on discrete space is used to enhance the alignment connection between speech and text and improve sampling quality; and 3) It uses an efficient text encoder to simplify the model's parameters and increase computational efficiency. The experimental results demonstrate that the approach proposed in this paper has outstanding speech synthesis quality and sampling speed while significantly reducing the resource consumption of diffusion model. The synthesized samples are available at \url{https://github.com/lawtherWu/DCTTS}
\end{abstract}
\section{Introduction}
\label{sec:intro}

Text-to-speech(TTS) aims to generate natural speech from input text, which has been the focus of the audio and speech processing research. Various TTS generative models are evolving. In terms of natural speech generation, Tacotron2 \cite{tacotron2}, FastSpeech2 \cite{fastspeech2} and TransformerTTS \cite{transtts} dominate the state of the art performance in Mean Opinion Score (MOS). The recently explored Diffusion Probabilistic Models (DPMs) \cite{dpms} serve as a powerful TTS generative backbone, achieving surprising results \cite{gradtts,fastdiff,prodiff}. These existing DPM-based speech synthesis works learn the connection between text and speech implicitly by adding a prior to the variational lower bound. Although such a method has excellent high fidelity and generalization, the resource consumption and slow inference of the diffusion model can not be ignored, which makes it difficult to use the diffusion model in practical usage scenarios.

These problems come from the high-dimensional raw speech features(e.g. spectrogram) and an excessive amount of diffusion steps. Although there are many improved methods to speed up the sampling of the diffusion model, they fail to address the underlying issues. Most work applies diffusion models to the continuous latent space, but much less studies apply them to the discrete space. Compressing data space is the fundamental way to solve the above problems.

This paper proposes a Discrete Diffusion Model with Contrastive Learning for Text-to-Speech Generation (DCTTS). This model will avoid the expensive cost of raw speech features prediction and generate natural speech with fewer diffusion steps. Specifically, the contributions of this paper are as follows: 

1. Propose a TTS diffusion model based on discrete space, compressing the data dimension of the diffusion model and increasing the computational efficiency; 

2. Propose contrastive learning  method based on discrete space to enhance the alignment connection between text and speech. We design the \emph{Text-wise Contrastive Learning Loss}(TCLL). The addition of TCLL enables the diffusion model to generate high fidelity speech samples with fewer diffusion steps; 

3. Introduce an efficient text encoder to further reduce the model parameters and computational consumption.

\section{Method}
\label{sec:method}

In this section, we present the details of the proposed DCTTS model whose architecture overview is shown in Figure 1. DCTTS consists of three parts, including a spectrogram VQ model, a text encoder and a discrete contrastive diffusion model. We first pre-train the spectrogram VQ model to compress the high-dimensional raw spectrogram into the discrete space. Then, the text encoder extracts text features from the input text.  The conditional diffusion model is used to predict the discrete token sequence conditioned on the text features. Finally, the discrete token sequence is decoded into the mel-sepctrogram by the spectrogram VQ decoder.

\begin{figure*}
    \centering
    \includegraphics[width=15cm]{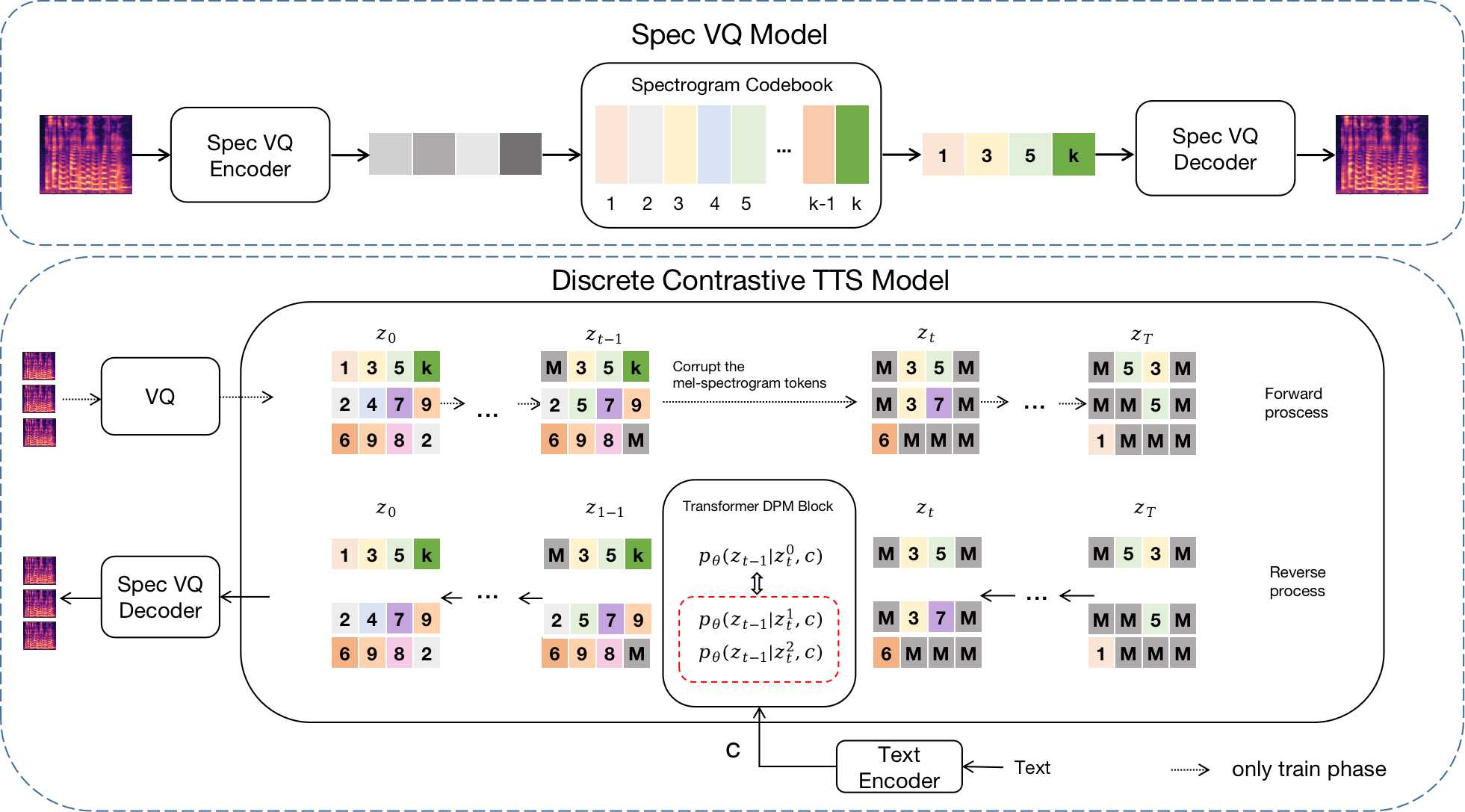}
    \caption{Overview of the proposed method. Our framework includes two major components: a Spectrogram VQ model (top) and a discrete conditional diffusion as generative model (bottom). The figure intuitively shows the forward process and reverse process of the discrete diffusion model. In the reverse process, we illustrate the proposed text-wise constrastive learning loss. }
    \label{Figure:model}
\end{figure*}

\subsection{Spectrogram VQ Model}
\label{ssec:specvq}
Reducing the computational cost of the diffusion model can be obtained by reducing the dimension of the data. Therefore, we propose to map the raw spectrogram to the discrete space. To achieve this, we introduce the Spectrogram VQ model\cite{vqgan,specvqgan} for vector quantization. As shown in the upper part of Figure 1, a spectrogram can be represented by a group of spectrogram tokens. Thus, the spectrogram generation transfers to predicting a sequence of discrete tokens. This method not only greatly reduces computational consumption, but also enhances the generalization of the TTS model\cite{vqtts}. Later, the diffusion model can be performed on the discrete space, largely avoiding the expensive raw spectrogram prediction.

The Spectrogram VQ Model is trained to approximate a spectrogram input using a compressed intermediate representation, retrieved from a discrete codebook. It consists of an encoder $E_{vq}$, a decoder $G$ and a codebook $\mathrm{Z}=\left\{\mathrm{z}_{\mathrm{k}}\right\}_{\mathrm{k}=1}^{\mathrm{K}} \in \mathbb{R}^{\mathrm{K} \times \mathrm{d}}$ containing a finite number of embedding vectors, where $K$ is the size of the codebook and $d$ is the dimension of codes. Given a spectrogram $s \in \mathbb{R}^{\mathrm{F} \times \mathrm{L}}$ where $F$ and $L$ represent frequency dimension and time dimension respectively, the input $s$ is firstly encoded into a small-scale representation $z \in E_{vq}(s) \in \mathbb{R}^{\mathrm{f} \times \mathrm{l}\times \mathrm{d}}$ where $f \times l$ represents the reduced frequency and time dimension. The $f \times l$ is usually much smaller than $F \times L$ . The $d$ represents the embedding dimension. Then, we use a spatial-wise quantizer $Q(.)$ which maps each spatial feature $z_{ij}$ ( for all (i, j) in (f, l) ) into its closest codebook entry $z_k$ to obtain a spatial collection of spectrogram tokens $z_q$.
\begin{equation} z_q=\mathrm{Q}\left(\mathrm{z}\right):=\left(\operatorname{argmin}_{\mathrm{z}_{\mathrm{k}} \in \mathrm{z}}\left\|\mathrm{z}_{\mathrm{ij}}-\mathrm{z}_{\mathrm{k}}\right\|_2^2\right) 
\end{equation}

Then the spectrogram can be faithfully reconstructed via the decoder i.e. $\hat{s}=G(z_q)$. To preserve the reconstruction quality when upsampled from a smaller-scale representation, we follow the setting of VQGAN \cite{vqgan}, which adds an additional discriminator module. The GAN structure enhances the reconstruction ability of the VQ model. In the next work of this paper, the parameters of the VQ model are frozen and will not be trained.

\subsection{Text Encoder}
\label{ssec:textencoder}
The text encoder aims to extract text representation from the input text. In previous works\cite{diffsound,vqdiff}, large pre-trained models are widely used as the text encoders, such as CLIP and BERT, which leads to huge model parameters and expensive computation. In this paper, we proposed an efficient text encoder.

The current EfficientSpeech\cite{efficientspeech} employs a fastspeech2-like network structure, which achieves efficient and high-quality speech synthesis. Inspired by the architecture of EfficientSpeech, our efficient text encoder is proposed, which consists of a phoneme encoder and an acoustic features extractor. The phoneme encoder extracts content features from the input phoneme obtained by g2p\cite{g2p}. The acoustic features extractor  predicts the Energy: $y_e$, Pitch: $y_p$ and Duration: $y_d$ from the content features. Instead of predicting the acoustic features in series \cite{fastspeech2}, the acoustic features extractor generates them in parallel which contributes to faster inference. The content features and acoustic features are concatenated together as the text features and input to the diffusion model.

\subsection{Discrete Diffusion Model with Contrastive Learning}
\label{ssec:DCtts}
Since the data dimensions are compressed using the VQ model, the process of inference can be achieved by training the diffusion model to predict discrete token sequences conditional on text features. This can greatly increase the speed of inference.The training of diffusion model includes the forward process and the reverse process, as shown in the lower part of Figure \ref{Figure:model}.

Given the text-spectrogram pair, we obtain the discrete spectrogram token sequence $z \in {\mathbb{Z}}^N$ with the pretrained Spectrogram VQ Model, where $N=f \times l$ represents the sequence length of tokens. Suppose the size of codebook is $K$, the spectrogram token $z_i$  at location $i$ takes the index that specifies the entries in the codebook, i.e. $z_i \in \{ 1,2,...,k\}$.

For forward phase, we corrupt the spectrogram tokens with Mask-and-replace diffusion strategy\cite{vqdiff}. The diffusion strategy follows a Markov chain and is defined as follows: each token has a probability of ${\gamma}_t$ to be masked by the \emph{[MASK]} token and has a probability of $K{\beta}_t$ to be resampled uniformly over all the $K$ categories, leaving the probability of ${\alpha}_t = 1 -k{\beta}_t -{\gamma}_t$ to be unchanged, whereas the \emph{[MASK]} tokens always keep fixed.

For reverse phase, the transformer DPM block is trained to predict and recover the corrupted token sequences based on condition input. The condition $c$ is the text features from the text encoder. The overall text-to-speech framework can be regarded as maximizing the conditional transition distribution $q\left(x \mid c\right)$. The network $p_\theta(x_{t-1}|x_t, y)$ is trained to estimate the posterior transition distribution $q_\theta(x_{t-1}|x_t, x_0)$. The optimization objective of this network is to minimize the variational lower bound\cite{vqdiff}.

For the text-to-speech task, the correspondence between text and speech is very important, which is directly related to the quality of synthesized speech. Misalignment between text and speech will lead to mispronunciation, such as repeating and dragging. Therefore, we hope that the text features maintain a good correspondence with the discrete spectrogram tokens. In this paper, we introduce the contrastive learning to help enhance the alignment connection.

We seek to enhance the connection between $c$ and the generated data $z_0$ by maximizing their mutual information, defined as $ I\left(z_0 ; c\right)=\sum_{z_0} p_\theta\left(z_0, c\right) \log \frac{p_\theta\left(z_0 \mid c\right)}{p_\theta\left(z_0\right)} $. We introduce a set of negative discrete spectrogram sequences $ Z'=\{ z_1,z_2,...,$ 
$z_N \} $, quantified from $N$ negative samples $X'=\{x_1,x_2,...x_N \}$. We define $f(z_0,c) = \frac{{p_{\theta}} (z_0|c)}{{p_{\theta}}(z_0)} $ and calculate the similarity of text features to control the degree of contrastive learning because similar texts should have similar spectrogram tokens. Our proposed \emph{Text-wise Contrastive Learning Loss} (TCLL) is:

\begin{scriptsize}
    \begin{equation}
    L_{TCLL}:=-\mathbb{E}\left [
    \log 
    \frac { f\left( z_0 , c \right)}
    { f\left(z_0,c\right) + \sum_{z_j \in Z'} \left\{ f(z_j,c) \left(1-sim(c,c_j)\right) \right\} }
    \right ]
\end{equation}
\end{scriptsize}
where $sim(c,c_j)$ indicates the Cosine similarity between $c$ and $c_j$. Optimization of this loss leads to maximization of $I(z_0; c)$. We add the TCLL to the optimization of the diffusion model by referring to \emph{Step-Wise Parallel Diffusion}\cite{cdcd}.

\section{EXPERIMENTAL AND RESULTS}
\label{sec:pagestyle}

\subsection{Dataset and Evaluation Metric}
\label{sec:dataset}
The dataset used for training is LJSpeech\cite{ljspeech} that is made of 13,100 audio clips with corresponding text transcripts. The training of model uses 12,588 clips while 512 clips are reserved for testing. The phoneme sequence is generated by the open-source tool g2p\cite{g2p} which convert  English grapheme to phoneme. The waveform is transformed into mel spectrogram with window and FFT lengths of 1,024, hop length of 256 and sampling rate of 22,050. The mel spectrogram has 80 channels. Montreal Force Alignment (MFA)  is used to obtain the target phoneme duration. Pitch and energy ground truth values are computed using STFT and WORLD vocoder\cite{world} respectively. Note that in order to exclude the influence of the vocoder, we uses the Griffin-Lim algorithm\cite{griffin} to convert the generated mel-spectrogram into waveform audio.

The DCTTS evaluation focuses on the generated speech quality. Moreover the evaluation gives priority to the number of parameters, amount of computations as measured by floating point operations (FLOPS), and sampling speed.

For Mean Opinion Score (MOS) estimation, we synthesized 20 sentences from the test split with each model. The assessors were asked to estimate the quality of synthesized speech on a nine-point Likert scale, the lowest and the highest scores being 1 point and 5 points with a step of 0.5 point.

The number of parameters refers to the amount of memory used in inference phase. GFLOPs reflects the number of floating point operations needed to complete an inference. GFLOPs increases with the text sequence length. In our experiment, GFLOPs is measured using 128 randomly sampled texts from the test split. The inputs to each model are same.

Sampling speed is usually measured in terms of Real-Time Factor(RTF is how many seconds it takes to generate one second of audio). However, The RTF leads to small fractional numbers that are less intuitive to interpret. We introduced mel spectrogram real-time-factor (mRTF) to measure the speed of DCTTS intuitively. mRTF is the number of seconds of speech divided by the mel spectrogram generation time\cite{efficientspeech}. 

\subsection{Implementation Details}
\label{sec:imple}

\textbf{Text Encoder}: The phoneme sequence $x_{phone} \in {\mathbb{R}}^{N\times d}$ is an embedding of the input phonemes, where $N$ is the sequence length and $d$ =128 is the embedding size. The phoneme encoder is made of 2 transformer blocks. Each block is made of a depth-wise separable convolution, a Self-Attention layer and a typical transformer FFN\cite{transformer} . In the FFN, we add an additional convolution layer and use the GeLU\cite{gelu} activation between two linear layers. Layer Normalization is applied after Self-Attention and FFN. Both Self-Attention and FFN use residual connection for fast convergence. The acoustic features extractor consists of 2 blocks. Each block includes a Convolution layer, Layer Normalization and a ReLU activation.
The prediction of Energy: $y_e$, Pitch: $y_p$ and Duration: $y_d$ are generated respectively by three extractors with same architecture.

\textbf{Spectrogram VQ Model}: In this study, we follow VQ-GAN, adopting similar network architecture for the VQ-VAE encoder $E_{vq}$, decoder $G$, and discriminator $D$. To preserve more timedimension information, we set a downsampling factor of 2 along the time axis, and a downsampling factor of 20 along the frequency axis. For the codebook $Z$, the dimension of each code vector $n_z$ is set as 128, and the codebook size $K$ is set as 128. The learning rate is fixed and determined as a product of a base learning rate, the number of GPUs used and the batch size. 

\textbf{Discrete contrastive diffusion model}: We built a 12-layer 8-head transformer with a dimension of 128 for the diffusion model. Each transformer block contains a full-context attention, a linear fusion layer to combine conditional features and a feed-forward network block. For the default setting, we set the total timesteps T = 100. The current timestep $t$ is added into the network with adaptive layer normalization operator. For the diffusion forward process, we linearly increase $\gamma_t$ and $\beta_t$ from 0 to 0.9 and 0.1, respectively.

\subsection{Comparison with Baseline}
\label{sec:comparison}
We compare the results with the baseline models \cite{tacotron2,fastspeech2,gradtts} which were evaluated based on official implementations. Note that we set the inference steps of Grad-TTS to 100(Grad-TTS-100), keeping the same inference steps as our DCTTS. 

Table \ref{tabel:Table1} shows the MOS evaluation metric as evaluated by 10 participants with high English listening comprehension. The synthesized speech samples are from the test split. Our results have great competitiveness in terms of audio quality, which indicates that our approach models the speech features  effectively . 

Not only that, our model is also very outstanding in efficiency, it has fewer parameters and GFlOPs that are used at inference phase. The effect of the small number of parameters and GFLOPS is faster mel spectrogram generation, reaching mRTF of 73.9 on a single NVIDIA 1080Ti GPU as shown in Table \ref{table:Table2}{}. The speed is more evident on an Intel CPU where DCTTS reaches mRTF of 17.6 which is 44.0× faster compared to Grad-TTS. For mel-spectrogram generation, Tacotron and Grad-TTS are unable to run with satisfactory mRTF on a single CPU.

\begin{table}[htb]
\centering
\caption{The comparison of parameters, GFlOPs and MOS.}
\begin{tabular}{l|l|l|l}
\hline
Model        & Params $\downarrow$ & GFlOPs $\downarrow$ & MOS $\uparrow$  \\ \hline
GT(mel)      & \quad -  & \quad -     & $3.86\pm 0.05$ \\
FastSpeech2 & 24.5m  & 15.87 & $3.58\pm 0.06$ \\
Tacotron2   & 28.2m  & 16.20 & $3.68\pm 0.05$ \\
Grad-TTS    & 14.8m  &  9.42  & $\bm{3.74\pm 0.07}$ \\
DCTTS       & \textbf{12.4m } &  \textbf{7.23}  & $3.64\pm 0.05$ \\ \hline
\end{tabular}

\label{tabel:Table1}
\end{table}

\begin{table}[htb]
\caption{The comparison of mRTF. The benchmarks were based on a single NVIDIA 1080Ti and a single Intel Xeon 2.50GHz.}
\resizebox{\linewidth}{!}{
\begin{tabular}{l|l|l|l|l}
\hline
Model       & \makecell{mRTF \\ GPU $\uparrow$} & \makecell{ES Relative \\Speed Up} & \makecell{mRTF \\ CPU $\uparrow$} & \makecell{ES Relative \\ Speed Up} \\ \hline
DCTTS       &    \textbf{73.9}  &    $\quad$ -  &     \textbf{17.6}   &    $\quad $-             \\
Grad-TTS    &    1.9   &   38.9 $\times $  &   0.4    &  44.0 $\times$    \\
Tacotron2   &    5.5   &   13.4  $\times$   &  1.3    &  13.5 $\times$        \\ \hline   
\end{tabular}
}

\label{table:Table2}
\end{table}

\subsection{Ablation Study}
\label{sec:ablation}

we conduct ablation experiments to study the contribution of the TCLL in our full model.

As shown in Table \ref{table:Table3}, by comparing the CMOS\cite{cmos} of DCTTS(w/o TCLL) and DCTTS, it is shown that the TCLL has a positive effect on the quality of generated speech.

\begin{table}[htb]
\centering
\caption{Comparison between DCTTS and DCTTS(w/o TCLL) in CMOS. TCLL refers to the Text-aware Contrastive Learning Loss}
\begin{tabular}{l|l}
\hline
Model            & CMOS $\uparrow$ \\ \hline
DCTTS            & 0    \\ \hline
DCTTS ( w/o TCLL) & -0.14     \\ \hline
\end{tabular}

\label{table:Table3}
\end{table}
\section{CONCLUSION}
\label{sec:conclusion}
By combining discrete diffusion model with contrastive learning, our method synthesizes natural speech with satisfactory inference speed. Besides, our model has fewer parameters and computational consumption. The proposed method pays inadequate attention to the emotional information in the input text. Future work will focus on extracting emotion features from text context and adding emotion features to speech generation.

\vfill\pagebreak
\bibliographystyle{IEEEbib}
\bibliography{refs}

\begin{thebibliography}{10}

\bibitem{tacotron2}
Jonathan Shen, Ruoming Pang, Ron~J Weiss, Mike Schuster, Navdeep Jaitly,
  Zongheng Yang, Zhifeng Chen, Yu~Zhang, Yuxuan Wang, Rj~Skerrv-Ryan, et~al.,
\newblock ``Natural tts synthesis by conditioning wavenet on mel spectrogram
  predictions,''
\newblock in {\em 2018 IEEE international conference on acoustics, speech and
  signal processing (ICASSP)}. IEEE, 2018, pp. 4779--4783.

\bibitem{fastspeech2}
Yi~Ren, Chenxu Hu, Xu~Tan, Tao Qin, Sheng Zhao, Zhou Zhao, and Tie-Yan Liu,
\newblock ``Fastspeech 2: Fast and high-quality end-to-end text to speech,''
\newblock {\em arXiv preprint arXiv:2006.04558}, 2020.

\bibitem{transtts}
Naihan Li, Shujie Liu, Yanqing Liu, Sheng Zhao, and Ming Liu,
\newblock ``Neural speech synthesis with transformer network,''
\newblock in {\em Proceedings of the AAAI conference on artificial
  intelligence}, 2019, vol.~33, pp. 6706--6713.

\bibitem{dpms}
Jascha Sohl-Dickstein, Eric Weiss, Niru Maheswaranathan, and Surya Ganguli,
\newblock ``Deep unsupervised learning using nonequilibrium thermodynamics,''
\newblock in {\em International conference on machine learning}. PMLR, 2015,
  pp. 2256--2265.

\bibitem{gradtts}
Vadim Popov, Ivan Vovk, Vladimir Gogoryan, Tasnima Sadekova, and Mikhail
  Kudinov,
\newblock ``Grad-tts: A diffusion probabilistic model for text-to-speech,''
\newblock in {\em International Conference on Machine Learning}. PMLR, 2021,
  pp. 8599--8608.

\bibitem{fastdiff}
Rongjie Huang, Max~WY Lam, Jun Wang, Dan Su, Dong Yu, Yi~Ren, and Zhou Zhao,
\newblock ``Fastdiff: A fast conditional diffusion model for high-quality
  speech synthesis,''
\newblock {\em arXiv preprint arXiv:2204.09934}, 2022.

\bibitem{prodiff}
Rongjie Huang, Zhou Zhao, Huadai Liu, Jinglin Liu, Chenye Cui, and Yi~Ren,
\newblock ``Prodiff: Progressive fast diffusion model for high-quality
  text-to-speech,''
\newblock in {\em Proceedings of the 30th ACM International Conference on
  Multimedia}, 2022, pp. 2595--2605.

\bibitem{vqgan}
Patrick Esser, Robin Rombach, and Bjorn Ommer,
\newblock ``Taming transformers for high-resolution image synthesis,''
\newblock in {\em Proceedings of the IEEE/CVF conference on computer vision and
  pattern recognition}, 2021, pp. 12873--12883.

\bibitem{specvqgan}
Vladimir Iashin and Esa Rahtu,
\newblock ``Taming visually guided sound generation,''
\newblock {\em arXiv preprint arXiv:2110.08791}, 2021.

\bibitem{vqtts}
Chenpeng Du, Yiwei Guo, Xie Chen, and Kai Yu,
\newblock ``Vqtts: High-fidelity text-to-speech synthesis with self-supervised
  vq acoustic feature,''
\newblock {\em arXiv preprint arXiv:2204.00768}, 2022.

\bibitem{diffsound}
Dongchao Yang, Jianwei Yu, Helin Wang, Wen Wang, Chao Weng, Yuexian Zou, and
  Dong Yu,
\newblock ``Diffsound: Discrete diffusion model for text-to-sound generation,''
\newblock {\em IEEE/ACM Transactions on Audio, Speech, and Language
  Processing}, 2023.

\bibitem{vqdiff}
Shuyang Gu, Dong Chen, Jianmin Bao, Fang Wen, Bo~Zhang, Dongdong Chen, Lu~Yuan,
  and Baining Guo,
\newblock ``Vector quantized diffusion model for text-to-image synthesis,''
\newblock in {\em Proceedings of the IEEE/CVF Conference on Computer Vision and
  Pattern Recognition}, 2022, pp. 10696--10706.

\bibitem{efficientspeech}
Rowel Atienza,
\newblock ``Efficientspeech: An on-device text to speech model,''
\newblock in {\em ICASSP 2023-2023 IEEE International Conference on Acoustics,
  Speech and Signal Processing (ICASSP)}. IEEE, 2023, pp. 1--5.

\bibitem{g2p}
K~Park and J~Kim,
\newblock ``g2pe,''
\newblock \url{https://github.com/ Kyubyong/g2p}, 2019.

\bibitem{cdcd}
Ye~Zhu, Yu~Wu, Kyle Olszewski, Jian Ren, Sergey Tulyakov, and Yan Yan,
\newblock ``Discrete contrastive diffusion for cross-modal music and image
  generation,''
\newblock in {\em The Eleventh International Conference on Learning
  Representations}, 2022.

\bibitem{ljspeech}
K~Ito and L~Johnson,
\newblock ``The lj speech dataset,''
\newblock \url{https: //keithito.com/LJ-Speech-Dataset/}, 2017.

\bibitem{world}
Masanori Morise, Hideki Kawahara, and Haruhiro Katayose,
\newblock ``Fast and reliable f0 estimation method based on the period
  extraction of vocal fold vibration of singing voice and speech,''
\newblock in {\em Audio Engineering Society Conference: 35th International
  Conference: Audio for Games}. Audio Engineering Society, 2009.

\bibitem{griffin}
Daniel Griffin and Jae Lim,
\newblock ``Signal estimation from modified short-time fourier transform,''
\newblock {\em IEEE Transactions on acoustics, speech, and signal processing},
  vol. 32, no. 2, pp. 236--243, 1984.

\bibitem{transformer}
Ashish Vaswani, Noam Shazeer, Niki Parmar, Jakob Uszkoreit, Llion Jones,
  Aidan~N Gomez, {\L}ukasz Kaiser, and Illia Polosukhin,
\newblock ``Attention is all you need,''
\newblock {\em Advances in neural information processing systems}, vol. 30,
  2017.

\bibitem{gelu}
Dan Hendrycks and Kevin Gimpel,
\newblock ``Gaussian error linear units (gelus),''
\newblock {\em arXiv preprint arXiv:1606.08415}, 2016.

\bibitem{cmos}
Philipos~C Loizou,
\newblock ``Speech quality assessment,''
\newblock in {\em Multimedia analysis, processing and communications}, pp.
  623--654. Springer, 2011.

\end{thebibliography}

\end{document}